\documentclass[12pt,english]{article}

\usepackage{geometry}
\usepackage{natbib}
\usepackage{url}
\makeatletter
\def\url@leostyle{%
    \def\UrlFont{\sf}}{\def\UrlFont{\small\ttfamily}}
\makeatother
\usepackage[font=small,labelfont=bf]{caption}

\usepackage[hidelinks]{hyperref}

\usepackage{authblk}

\usepackage{graphicx}
\usepackage{babel}

\usepackage{amsmath}
\usepackage{amsfonts}

\usepackage{framed}


\numberwithin{equation}{section}

\begin{document}

\title{On the Significance of the Gottesman-Knill Theorem\thanks{\textbf{Note:} this is the submitted version of
    an article that is to appear in \emph{The British Journal
    for the Philosophy of Science}, published by \emph{Oxford
    University Press.} Changes have been made to this article since it
    was submitted for publication. These changes, while substantive, are
    clarificatory in nature and do not modify the main claim
    or overall structure of the argument of the paper. When citing this
    paper, please refer to the published version.
  }$\mbox{ }^,$\thanks{I am very significantly indebted, first and
    foremost, to Wayne Myrvold for our many and fruitful discussions
    on the topic of this paper, and for commenting on my earlier
    drafts. I also benefited substantially from my discussions on the
    topic with William Demopoulos, Lucas Dunlap, and Samuel
    Fletcher. I am very grateful, additionally, to Jeffrey Barrett,
    Armond Duwell, and Ryan Samaroo, who were kind enough to take the
    time to read my earlier drafts and to provide me with constructive
    comments and criticisms. Finally I am indebted to the Alexander
    Humboldt Foundation, whose financial support has made this project
    possible.
  }
}

\author[]{Michael E. Cuffaro}
\affil[]{Munich Center for Mathematical Philosophy, Ludwig
  Maximilians Universit\"at M\"unchen}
\date{}

\maketitle

\thispagestyle{empty}

\begin{abstract}
{
\footnotesize
According to the Gottesman-Knill theorem, quantum algorithms which
utilise only the operations belonging to a certain restricted set are
efficiently simulable classically. Since some of the operations in
this set generate entangled states, it is commonly concluded that
entanglement is insufficient to enable quantum computers to outperform
classical computers. I argue in this paper that this conclusion is
misleading. First, the statement of the theorem (that the particular
set of quantum operations in question can be simulated using a
classical computer) is, on reflection, already evident when we
consider Bell's and related inequalities in the context of a
discussion of computational machines. This, in turn, helps us to
understand that the appropriate conclusion to draw from the
Gottesman-Knill theorem is not that entanglement is insufficient to
enable a quantum performance advantage, but rather that if we limit
ourselves to the operations referred to in the Gottesman-Knill
theorem, we will not have used the resources provided by an entangled
quantum system to their full potential.
}
\end{abstract}

\section{Introduction}

Toil in the field of quantum computation promises a bountiful harvest,
both to the pragmatic-minded researcher seeking to develop new and
efficient solutions to practical problems of immediate and transparent
significance, as well as to those of us moved more by philosophical
concerns: we who toil in the mud and black earth, ever desirous of
those remote and yet more profound insights at the root of scientific
inquiry. Some of us have seen in quantum computation the promise of a
solution to the interpretational debates which have characterised the
foundations and philosophy of quantum mechanics since its
inception. Some of us have seen the prospects for a deeper
understanding of the nature of computation as such. Others have seen
quantum computation as potentially illuminating our understanding of
the nature and capacities of the human mind.\footnote{For an
  example of a claim to the effect that quantum computation is
  capable of informing interpretational debates in quantum
  mechanics, see \citet{deutsch1997}, and for further discussion of
  this, see \citet[]{wallace2012}. Discussions of the relevance of
  quantum computation to mathematics and computer science can be
  found in \citet{deutsch2000}, \citet{hagar2007b},
  \citet[]{aaronson2013}, and \citet{timpson2013}. For claims of
  relevance to the philosophy of mind, see, e.g.,
  \citet[]{hameroff1998}.}

An arguably more modest position \citep[see,
  e.g.,][]{aaronson2013,timpson2013} regarding the philosophical
interest of quantum computation (and related fields like quantum
information), is that its study contributes to our understanding of
the foundations of quantum mechanics and computer science mainly by
offering us different perspectives on old foundational
questions---fresh opportunities, that is, to reconsider just what we
mean in asking these questions. One of my goals in this paper is to
provide such a different perspective, on the Bell inequalities in
particular. Specifically I will be arguing that the significance we
attach to Bell's and related inequalities is in part informed by the
context of discussion. It is informed, that is, by a number of what I
will call `plausibility constraints' associated with a given context,
whose role is to rule out certain `loopholes' to the inequalities in
that context. Consequently, the kind of local hidden variables
descriptions that we deem plausible in the context of a discussion of
the sorts of systems buildable \emph{by us}, is importantly different
from what we deem plausible in the context of a discussion of the
natural world as it exists apart from us. And yet in both cases there
is still an important sense in which the Bell inequalities should be
taken as answering the same question: the question, that is,
concerning the boundary between classical and quantum description.

This is my first goal. My ultimate goal, however (which is informed by
the first), is to clarify a narrower issue. Specifically it is to
clarify the discussion surrounding the claim that the presence of a
quantum system in a pure entangled state is sufficient to allow a
quantum computer to realise a computational advantage (or `quantum
speedup') over a classical computer.\footnote{I will be focusing
  exclusively on the computational capabilities associated with pure
  states in this paper. A system in a mixed entangled state can be
  thought of as being in the presence of `noise', strong enough, in
  some cases, to prevent the system from being capable of realising
  more than a very small speedup \citep[see][]{linden2001}. Our
  concern, however, is mainly with the issue of whether even an
  entangled system not in the presence of any noise whatsoever
  (i.e., a pure state) is sufficient to enable speedup. I discuss this
  issue again in n. \ref{fn:roughwaves} below.} Call this claim the
``sufficiency of entanglement thesis.'' In the literature on quantum
computation, one often encounters the statement that the sufficiency
of entanglement thesis is false. Motivating those who would deny the
sufficiency of entanglement thesis is the Gottesman-Knill theorem
\citep{gottesman1999}. According to this theorem, any quantum
algorithm which exclusively utilises the elements of a particular
restricted set of quantum operations can be re-expressed using an
alternative (i.e., the `stabiliser') formalism which shows us how that
algorithm can be efficiently simulated by classical means. Since some
of the algorithms which exclusively utilise operations from this set
involve the generation of pure entangled states, it seems that
entanglement cannot therefore be sufficient to enable one to achieve a
quantum speedup. Thus, \citeauthor[]{datta2005}, for instance, write:
``the Gottesman-Knill theorem ... demonstrates that global
entanglement is far from sufficient for exponential speedup.''
\citeyearpar[1]{datta2005}. \citeauthor[]{nielsenChuang2000} likewise
write:

\begin{quote}
Consider that interesting quantum information processing tasks ... can
therefore be efficiently simulated on a classical computer, by the
Gottesman-Knill theorem. Moreover, we will see shortly that a wide
variety of quantum error-correcting codes can be described within the
stabilizer formalism. There is much more to quantum computation than
just the power bestowed by quantum entanglement!
\citeyearpar[464]{nielsenChuang2000}.
\end{quote}

We will examine a more detailed and explicit statement of this
position by \citet[]{jozsa2003} in \textsection
\ref{sec:siggkthm}. For now let me simply say that, in the quotations
cited above, it is not immediately obvious exactly what is meant when it
is said that entanglement is ``far from sufficient'' for speedup, or
that ``there is much more to quantum computation'' than
entanglement. Part of my goal here, therefore, is to distinguish and
clarify the different senses in which such statements can be
taken. There is of course one sense in which the Gottesman-Knill
theorem shows, conclusively, that the sufficiency of entanglement
thesis is false: clearly, the mere presence of a pure entangled state
is insufficient to realise a quantum computational speedup. However
the sufficiency of entanglement thesis can also be taken in a second,
more interesting sense. One might take the thesis, that is, as
claiming that quantum entanglement, by itself, is sufficient \emph{to
  enable}, or \emph{make possible}, quantum computational speedup;
i.e., that \emph{no other physical resources are needed} to make
quantum speedup possible if one begins with a system in a pure
entangled state. This claim, or so I will argue, is \emph{not} proved
false by the Gottesman-Knill theorem. What the Gottesman-Knill theorem
shows us, rather, is only that if we limit ourselves to the
Gottesman-Knill operations, we will not have used the entanglement
with which we have been provided to its full potential. For the
Gottesman-Knill operations, I will argue, are just those operations
whose associated statistics (even when they involve entangled states)
are reproducible in a local hidden variables theory that we would
grant as plausible in the computational context. Indeed we do not need
the Gottesman-Knill theorem to tell us this, for as I will argue, it
is already evident when we consider the Bell inequalities.\footnote{If
  this is correct, then why do I not say that the sufficiency of
  entanglement thesis is \emph{true}, rather than merely that it is
  not proved false? Explaining quantum speedup involves more than
  showing why a classical computer is incapable of efficiently
  simulating the evolution of a quantum computer. It involves, in
  addition, showing why a classical computer cannot, \emph{by
  whatever means}, solve a particular problem efficiently which
  happens to be in the complexity class BQP (the ``bounded error
  quantum polynomial time'' class). These two questions are
  obviously related, and it is not at all implausible to think that
  the explanation will be the same in both cases. However this is by
  no means automatic.
}

The paper will go as follows: I will introduce the Gottesman-Knill
theorem and motivate the assertion that the sufficiency of
entanglement thesis is in some sense false in \textsection
\ref{sec:gkthm}. Then in \textsection \ref{sec:siggkthm} I will begin
by considering a detailed and explicit version of this assertion due
to \citet[]{jozsa2003}. In the remainder of \textsection
\ref{sec:siggkthm} I will try to clarify what I take to be the real
lesson of the Gottesman-Knill theorem: that---and this is in an
important sense nothing new---there are some statistics associated with
entangled states which admit of a local hidden variables
description.\footnote{I say that the Gottesman-Knill theorem tells
  us nothing new. While this is true in an important sense, I should
  not be taken here as denying that the theorem is of value. Despite
  the fact that in one sense it is, as we will see, no more than a
  corollary of Bell's theorem, it is nevertheless a profoundly
  illuminating result in that it encourages us to look at Bell's and
  other inequalities with fresh eyes and in different contexts.} In
\textsection \ref{sec:explic} I will argue that we should consider
these descriptions to be plausible in the context of a discussion of
classical and quantum computation. Finally in \textsection
\ref{sec:suff} I will consider the consequences of this for our
understanding of the sufficiency of entanglement thesis. I will argue
that if one intends by the claim that entanglement is sufficient to
enable quantum speedup, that no further physical resources are
required (the claim denied by \citeauthor{jozsa2003}), then this claim
is not shown false by the Gottesman-Knill theorem.

\section{The Gottesman-Knill theorem}
\label{sec:gkthm}

Call an operator $A$ a \emph{stabiliser} of the state $| \psi \rangle$
if:\footnote{In the following exegesis of the Gottesman-Knill theorem
  I have drawn substantially from \citet[]{gottesman1999} and from
  \citet[]{nielsenChuang2000}.}
\begin{align}
A| \psi \rangle = | \psi \rangle.
\end{align}
For instance, consider the Bell state of two qubits:\footnote{A qubit
  is the basic unit of quantum information, analogous to a classical
  bit. It can be physically realised by any two-level quantum
  mechanical system.}
\begin{align*}
| \Phi^+ \rangle = \frac{1}{\sqrt 2}(| 0 \rangle \otimes | 0 \rangle +
| 1 \rangle \otimes | 1 \rangle).
\end{align*}
For this state we have
\begin{align*}
(X \otimes X)| \Phi^+ \rangle & = \frac{1}{\sqrt 2}(| 1 \rangle
  \otimes | 1 \rangle + | 0 \rangle \otimes | 0 \rangle) \nonumber \\
  & = \frac{1}{\sqrt 2}(| 0 \rangle \otimes | 0 \rangle + | 1 \rangle
  \otimes | 1 \rangle) = | \Phi^+ \rangle, \\
(Z \otimes Z)| \Phi^+ \rangle & = \frac{1}{\sqrt 2}(| 0 \rangle
  \otimes | 0 \rangle + (-| 1 \rangle \otimes -| 1 \rangle)
  \nonumber \\
  & = \frac{1}{\sqrt 2}(| 0 \rangle \otimes | 0 \rangle + | 1 \rangle
  \otimes | 1 \rangle) = | \Phi^+ \rangle.
\end{align*}
$X \otimes X$ and $Z \otimes Z$ are thus both stabilisers of the state
$| \Phi^+ \rangle$. Here, $X$ and $Z$ are the Pauli operators:
\begin{align*}
  X \equiv
  \left(
  \begin{matrix}
  0 & 1 \\
  1 & 0
  \end{matrix}
  \right), &
  \quad Z \equiv
  \left(
  \begin{matrix}
  1 & 0 \\
  0 & -1
  \end{matrix}
  \right).
\end{align*}
The remaining Pauli operators, $I$ (the identity operator) and $Y$,
are defined as:
\begin{align*}
  I \equiv
  \left(
  \begin{matrix}
  1 & 0 \\
  0 & 1
  \end{matrix}
  \right), &
  \quad Y \equiv
  \left(
  \begin{matrix}
  0 & -i \\
  i & 0
  \end{matrix}
  \right).
\end{align*}
The set $P_n$ of $n$-fold tensor products of Pauli operators
(plus those multiplied by $\alpha \in \{\pm 1, \pm i\}$) forms a
group of operators, called a \emph{Pauli Group}, which is closed under
matrix multiplication. For example, for $n = 2$: $P_2 \equiv \{\alpha
I \otimes I$, $\alpha I \otimes X$, $\alpha I \otimes Y$, $\alpha I
\otimes Z$, $\alpha X \otimes I$, $\alpha X \otimes X$, $\alpha X
\otimes Y$, $... \}$.\footnote{The Pauli operators $I$, $X$, $Y$, $Z$
  are rare in that they are both unitary and Hermitian operators (it
  is because of the latter, of course, that they are also called the
  Pauli \emph{observables}). When we generalise these operators to
  allow multiples $\alpha = \{\pm 1, \pm i\}$, however, this is
  sometimes no longer the case. For example, the operators $iX$ and
  $-iX$, though unitary, are not Hermitian (they are anti-Hermitian).}

Call the set $V_S$ of states that are stabilised by every element in
$S$, where $S$ is some group of operators closed under matrix
multiplication, the \emph{vector space stabilised by} $S$. Consider a
state $| \psi \rangle \in V_S$. For any $s \in S$ and any unitary
operation $U$, we have
\begin{align}
U| \psi \rangle = Us| \psi \rangle = UsU^\dagger U| \psi \rangle,
\end{align}
where the last equality follows from the definition of a unitary
operator. Thus $UsU^\dagger$ stabilises $U| \psi \rangle$ and the
vector space $UV_S$ is stabilised by the group $USU^\dagger \equiv
\{UsU^\dagger|s \in S\}$. Consider, for instance, the state $| 0
\rangle$, stabilised by the $Z$ operator. To determine the stabiliser
of this state after it has been subjected to the (unitary)
Hadamard\footnote{The $H$ or Hadamard operator takes $| 0 \rangle$ to
  $(| 0 \rangle + | 1 \rangle)/\sqrt 2 \equiv | + \rangle$ and
  $| 1 \rangle$ to $(| 0 \rangle - | 1 \rangle)/\sqrt 2 \equiv | -
  \rangle$.} transformation $H| 0 \rangle = | + \rangle$ we simply
compute $HZH^\dagger$. Thus the stabiliser of $| + \rangle$ is $X$.

Now let $s_1, ..., s_m$ be elements of $S$. $s_1, ..., s_m$ are said
to \emph{generate} the group $S$ if every element of $S$ can be
written as a product of elements from $s_1, ..., s_m$. For instance,
the reader can verify that the subgroup, $A$, of $P_3$, defined by $A
\equiv \{I \otimes I \otimes I, Z \otimes Z \otimes I, I \otimes Z
\otimes Z, Z \otimes I \otimes Z\}$ can be generated by the
elements $\{Z \otimes Z \otimes I, I \otimes Z \otimes Z\}$
\citep[\textsection 10.5.1]{nielsenChuang2000}. We may thus
alternately express $A$ in terms of its generators as follows: $A =
\langle Z \otimes Z \otimes I, I \otimes Z \otimes Z \rangle$.

In order to compute the action of a unitary operator on a group $S$ it
suffices to compute the action of the unitary operator on the
generators of $S$. For instance, $| 0 \rangle^{\otimes n}$ is the
unique state stabilised by $\langle Z_1, Z_2, ..., Z_n\rangle$ (where
the latter expression is a shorthand form of $\langle Z \otimes
I^{\otimes n-1}, I \otimes Z \otimes I^{\otimes n-2}, ...,I^{\otimes
  n-1} \otimes Z\rangle$). Consequently, the stabiliser of the state
$H^{\otimes n}| 0 \rangle^{\otimes n}$ is $\langle X_1, X_2, ...,
X_n\rangle$. Note that this state, expressed in the standard state
vector formalism:\footnote{Note that from now on $|\alpha\beta
  \rangle$ and $| \alpha \rangle| \beta \rangle$ should be understood
  as shorthand forms of $| \alpha\rangle\otimes| \beta
  \rangle$. Additionally: $| \alpha \rangle_1 \otimes | \alpha
  \rangle_2 \otimes ... \otimes | \alpha \rangle_n \equiv | \alpha^n
  \rangle \equiv | \alpha \rangle^n \equiv | \alpha \rangle^{\otimes
    n}.$}
\begin{align}
\label{eqn:multihadprod}
H^{\otimes n}| 0 \rangle^{\otimes n} = \frac{1}{2^{n/2}}(| 0
\rangle + | 1 \rangle)_1 (| 0 \rangle + | 1 \rangle)_2 \dots (| 0
\rangle + | 1 \rangle)_n \\
\label{eqn:multihadsup}
= \frac{1}{2^{n/2}}(| 00 \dots 00 \rangle + | 00 \dots 01
\rangle + \dots + | 11 \dots 10 \rangle + | 11 \dots 11 \rangle,
\end{align}
specifies $2^n$ different amplitudes. Contrast this with the
stabiliser description of the state in terms of its generators
$\langle X_1, X_2, ..., X_n\rangle$, which is linear in $n$ and thus
capable of an efficient classical representation.\footnote{A basic
  distinction, in computational complexity theory, is between those
  computational problems that are amenable to an \emph{efficient}
  solution in terms of time and/or space resources, and those that
  are not. Easy (or `tractable', `feasible', `efficiently solvable',
  etc.) problems are those for which solutions exist which involve
  resources bounded by a polynomial in the input size, $n$. Hard
  problems are those which are not easy, i.e., they are those whose
  solution requires resources that are `exponential' in $n$, i.e.,
  that grow faster than any polynomial in $n$
  \citep[p. 139]{nielsenChuang2000}. Note that the term
  `exponential' is being used rather loosely here. Functions such as
  $n^{\log n}$ are called `exponential' but do not grow as fast as a
  true exponential such as $2^{n}$.}

Using the stabiliser formalism, it can be shown that all, as well as
all combinations, of the following operations are capable of an
efficient classical representation. (i) The \emph{Clifford group} of
gates; i.e., those unitary transformations which map elements of the
Pauli group to other elements of the Pauli group.\footnote{A quantum
  `gate' is just a unitary transformation. In a quantum
  computational circuit it plays a role analogous to a (reversible)
  logic gate in a classical circuit.} These are the Pauli ($I$, $X$,
$Y$, $Z$) gates, the Hadamard gate, the Phase gate (a $\pi/2$ rotation
of the Bloch sphere\footnote{The Bloch sphere is a geometrical
  representation of the state space of a single qubit. States on the
  surface of the sphere represent pure states, while those in the
  interior represent mixed states \citep[see][]{nielsenChuang2000}.}
about the $\hat{z}$-axis), and the controlled not (``CNOT'')
gate.\footnote{The CNOT or controlled-not gate takes two qubits $| s
  \rangle| t \rangle$ to $| s \rangle| t \oplus s \rangle$, where $|
  s \rangle$ is the control, $| t \rangle$ the target qubit, and
  $\oplus$ is addition modulo 2 (i.e., `exclusive-or'). Intuitively,
  the control qubit determines whether or not to apply a bit-flip
  operation (i.e., a NOT or $X$ operation) to the target
  qubit.}$^,$\footnote{Note that the Hadamard, Phase, and CNOT gates
  by themselves suffice to generate the Clifford Group.} (ii)
Clifford group gates conditioned on classical bits (indicating, e.g.,
the results of previous measurements). (iii) State preparation in the
computational (i.e., $\{| 0 \rangle, | 1 \rangle\}$) basis. (iv)
Measurements of observables in the Pauli group. This is the content of
the \emph{Gottesman-Knill theorem} \citep[\textsection
  10.5.4]{nielsenChuang2000}.

What is especially notable about this theorem from the point of view
of our discussion is that some of the states which may be realised
through the operations in this set are actually entangled states. In
particular, by combining a Hadamard and a CNOT gate, one can generate
any one of the Bell states (which one is generated depends on the
value assigned to the input qubits); i.e.,
\begin{align}
\label{eqn:phiplus}
  | 0 \rangle| 0 \rangle \xrightarrow{H \otimes I} \frac{| 0 \rangle| 0
    \rangle + | 1 \rangle| 0 \rangle}{\sqrt 2}
  \xrightarrow{\mbox{\scriptsize{CNOT}}} \frac{| 0 \rangle| 0 \rangle +
    | 1 \rangle| 1 \rangle}{\sqrt 2} = | \Phi^+ \rangle, \\
\label{eqn:psiplus}
  | 0 \rangle| 1 \rangle \xrightarrow{H \otimes I} \frac{| 0 \rangle| 1
    \rangle + | 1 \rangle| 1 \rangle}{\sqrt 2}
  \xrightarrow{\mbox{\scriptsize{CNOT}}} \frac{| 0 \rangle| 1 \rangle +
    | 1 \rangle| 0 \rangle}{\sqrt 2} = | \Psi^+ \rangle, \\
\label{eqn:phiminus}
  | 1 \rangle| 0 \rangle \xrightarrow{H \otimes I} \frac{| 0 \rangle| 0
    \rangle - | 1 \rangle| 0 \rangle}{\sqrt 2}
  \xrightarrow{\mbox{\scriptsize{CNOT}}} \frac{| 0 \rangle| 0 \rangle
    - | 1 \rangle| 1 \rangle}{\sqrt 2} = | \Phi^- \rangle, \\
\label{eqn:psiminus}
  | 1 \rangle| 1 \rangle \xrightarrow{H \otimes I} \frac{| 0 \rangle| 1
    \rangle - | 1 \rangle| 1 \rangle}{\sqrt 2}
  \xrightarrow{\mbox{\scriptsize{CNOT}}} \frac{| 0 \rangle| 1 \rangle
    - | 1 \rangle| 0 \rangle}{\sqrt 2} = | \Psi^- \rangle.
\end{align}
In fact many quantum algorithms utilise just such a combination of
gates (e.g., teleportation; see
\citealt[\textsection 1.3.7]{nielsenChuang2000}). Now recall that the
sufficiency of entanglement thesis is the claim that the presence of a
quantum system in a pure entangled state is sufficient to allow a
quantum computer to realise quantum speedup. If all of the operations
from this set are efficiently classically simulable, however, then it
appears as though the sufficiency of entanglement thesis must be
false (at least in some sense), for evidently there are quantum
algorithms utilising pure entangled states that are efficiently
simulable classically.\footnote{It is worth noting that
  \citet[473]{steane2003} uses the Gottesman-Knill theorem as the
  basis for an objection to the many worlds explanation of quantum
  computation (the many worlds explanation seems, at least at first
  blush, to depend crucially on the fact that quantum algorithms are
  expressible in the state vector formalism). His own view is that it
  is entanglement which explains quantum speedup. It should be clear,
  however, that the Gottesman-Knill theorem is just as much a prima
  facie problem (if not more so) for Steane's view as it is for many
  worlds theorists.}

\section{The significance of the Gottesman-Knill theorem}
\label{sec:siggkthm}

Reflecting on this circumstance in their influential
\citeyearpar{jozsa2003} article (in a section entitled: \emph{Is
  entanglement a key resource for computational power?}),
\citeauthor[]{jozsa2003} write:

\begin{quote}
Recall that the significance of entanglement for pure-state
computations is derived from the fact that unentangled pure states
... of $n$ qubits have a description involving poly($n$) parameters
(in contrast to $O(2^n)$ parameters for a general pure state). But
this special property of unentangled states (of having a `small'
descriptions [\emph{sic.}]) is contingent on a particular mathematical
description, as amplitudes in the computational basis. If we were to
adopt some other choice of mathematical description for quantum states
(and their evolution), then, although it will be mathematically
equivalent to the amplitude description, there will be a different
class of states which will now have a polynomially sized description;
i.e. two formulations of a theory which are mathematically equivalent
(and hence equally logically valid) need not have their corresponding
mathematical descriptions of elements of the theory being
[\emph{sic.}] interconvertible by a \emph{polynomially bounded}
computation. With this in mind we see that the significance of
entanglement as a resource for quantum computation is not an
\emph{intrinsic} property of quantum physics \emph{itself}, but is
tied to a particular additional (arbitrary) choice of mathematical
formalism for the theory. ... An explicit example of an alternative
formalism and its implications for the power of quantum computation is
provided by the so-called stabilizer formalism and the Gottesman-Knill
theorem ... Thus, in a fundamental sense, the power of quantum
computation over classical computation ought to be derived
simultaneously from \emph{all} possible classical mathematical
formalisms for representing quantum theory, not any single such
formalism and associated quality (such as entanglement),
... \citep[2029-2030][emphasis in original]{jozsa2003}.
\end{quote}

That two equivalent mathematical representations of the same physical
object can have ineliminably vastly differently sized descriptions
\emph{depending only on the formalism used} is a profoundly
counter-intuitive notion. This is not to say that it cannot be
correct, of course. But rather than trying to make sense of this
notion, let us consider whether some other more satisfying explanation
can be given. It is easy to see, first of all, that even solely within
the amplitude formalism, one and the same system can admit of either a
large or a small description. We have seen an example of this
already. The state that results from applying $H^{\otimes n}$ to a
system in the state $| 0 \rangle^{\otimes n}$ can be described as a
superposition of $2^n$ states, as in Eq. \eqref{eqn:multihadsup}. It
can also be described as a product of $n$ states, as in
Eq. \eqref{eqn:multihadprod}\textemdash an exponentially smaller
description. Indeed we can do much better than this, and can make do
with one of the even more compact expressions:
\begin{align}
\label{eqn:multihadexporsum}
\frac{1}{2^{n/2}}(| 0 \rangle + | 1 \rangle)^n\mbox{ }, \qquad\quad
\frac{1}{2^{n/2}}\sum_x^{2^n-1}| x \rangle.
\end{align}
All of these descriptions are equivalent. In this case, however, there
is no mystery as to why. It is facts about the underlying systems
being described which make these differently sized descriptions
possible. For instance, the fact that the properties of each
individual subsystem are maximally specifiable makes it possible to
represent the superposition \eqref{eqn:multihadsup} as the product
state \eqref{eqn:multihadprod}. And since in this particular case each
subsystem is in an identical state, we do not really need to single
out any one of them, and thus we can use one of the descriptions given
in \eqref{eqn:multihadexporsum}.

This is not true in general. It is a quantum mechanical fact that
subsystems of entangled systems are not maximally specifiable (i.e.,
their states are never pure). Thus entangled quantum systems cannot be
given a pure product state representation. Descriptions of entangled
states and of the transitions to and from them cannot therefore be
compressed in the same way that \eqref{eqn:multihadsup} is
compressible into \eqref{eqn:multihadprod}. At least this is true in
the standard amplitude formalism. Strangely, if we move from the
amplitude to the stabiliser formalism it seems as though it \emph{is}
possible, somehow, to give more compact descriptions of these states
and their transitions, despite the quantum mechanical fact just
mentioned.

However let us persist, for a little while longer at least, in our
conviction that it is facts about the underlying system that is
subjected to the Gottesman-Knill transformations, and not facts about
the formalism we use to describe it per se, which makes this
possible. If we persist in this, then we are led naturally to the
conclusion that systems subjected to just these transformations are
somehow not using the entanglement available to them effectively, that
somehow the information that would be required to specify them if they
were using it effectively is not needed and can be omitted. This
statement is still quite vague. Let us see if we can explicate it more
precisely.

The Gottesman-Knill theorem tells us that a certain set of quantum
operations can be efficiently simulated with a classical computer
\citep[464]{nielsenChuang2000}. Let us then begin by explicating the
notion of ``classical computer simulation.'' What does it mean to
provide a classical computer simulation of a quantum system? Perhaps
most essentially, it means that the computer doing the simulating can
be given a \emph{classical description}. What, now, does this mean?
Well for one, it means that a complete description of the system,
which in general can consist of many sub-components, is separable into
complete descriptions of those individual sub-components. Furthermore,
it means that our descriptions of the interactions between those
sub-components are always constrained by classical physical
laws.\footnote{With Bohr, I take these to include the laws of
  special and general relativity.} For instance, the speed by which
these interactions propagate will be constrained by the speed of
light. Importantly, the behaviour associated with classical systems
can always be described (in principle) in terms of causes and effects,
and correlations between effects can always be described in terms of
common causes. Thus a classical computer is called classical because
it can be described as a classical physical system. Its components,
describable independently of one another, interact in a locally
causal, spatiotemporally continuous, manner. They can never be
interpreted as violating any of the laws and principles of special
relativity or of classical mechanics. Nor can the correlations between
spatially distant parts of the machine ever be interpreted as
violating the Bell inequalities. This last statement is true no matter
how large the machine is, and no matter how far apart the different
nodes of the computer (e.g., in a distributed setting) are. All such
correlations are always factorisable and they can always in principle
be described in terms of classical common causes; i.e., common causes
that are located within the prior light cone of these spatially
distant locations.

Now imagine a quantum computer performing some series of
Gottesman-Knill operations, and imagine a classical computer which
efficiently simulates those operations. If I were to now provide you
with a specification of the classical computer---its source code, for
instance, or better: some hardware-level description---then in doing
this I would be providing you with an alternative classical,
i.e. locally causal, description of what the quantum computer is doing
when it is performing those operations. In other words, I would be
providing you with a ``local hidden variables theory,'' so to speak,
to reproduce the quantum computer's observable behaviour. This
deserves emphasis: any classical computer which we program to keep
track of the stabiliser description of a quantum system as it evolves
through time is describable in a locally causal way. This is just what
it means for the computer to be classical. To describe the classical
computer's operation as it simulates the quantum computer, therefore,
is to provide a local hidden variables theory for the latter.

Indeed, contrary to \citeauthor[]{jozsa2003}, there is no need to
refer to the stabiliser formalism in order to show that one can
describe the Gottesman-Knill operations in terms of such local hidden
variables theories. We can show this to be true by referring to the
standard \emph{amplitude} formalism as well. This is evident if we
consider the implications of the Bell inequalities. Recall that the
Gottesman-Knill operations consist specifically of the Clifford group
of transformations (possibly conditioned on classical bits) which map
the Pauli group into itself, measurements of observables in the Pauli
group, and state preparation in the computational basis. In Appendix
\ref{sec:appendixBellNoCc} it is shown (informally) that the combined
effect of any of these operations, for any subsystem of the system to
which they are applied, is equivalent to the measurement of one of the
Pauli observables $\pm X$, $\pm Y$, $\pm Z$ on an eigenstate of a
Pauli observable. Now since the Pauli observables (disregarding the
trivial transformation $I$) represent $\pi$ rotations of the Bloch
sphere about the $x$, $y$, and $z$ axes, the respective orientations
of different ends of an experimental apparatus set up to conduct an
experiment involving Pauli observables on a combined system will never
differ by anything other than an angle proportional to $\pi/2$. These,
however, are precisely the orientations for which Bell showed it
possible to provide a local hidden variables theory to reproduce the
statistics associated with the singlet state
\citep[]{bell1964}. Bell's technique is straightforwardly extendable,
moreover, to the other Bell states and to other bipartite entangled
combinations of eigenstates of Pauli operators (see Appendix
\ref{sec:appendixBellNoCc}).\footnote{In fact, as we will discuss
  later, there is a yet simpler way (involving a small amount of
  classical communication) to recover the statistics associated with
  these states (see Appendix \ref{sec:appendixBellCc}).}

Thus far we have considered only bipartite systems. Yet it is also
possible to provide local hidden variables descriptions of the
statistics associated with Pauli measurements on general $n$-partite
systems, and for such cases things are more interesting. For instance,
in the tripartite GHZ (a.k.a. ``cat'') state: $(| 000 \rangle + | 111
\rangle)/\sqrt 2$, it is well known that it is impossible to assign
noncontextual values to the $x$ and $y$ components of each particle's
spin in a way that recovers the quantum mechanical predictions
associated with (simultaneous) joint measurements of Pauli observables
on the combined system \citep[]{mermin1990}. One can nevertheless
produce a local hidden variables description of these measurement
statistics if the nodes of the classical computer one is describing
are able to communicate with one another during the process. The
communication of only one classical bit, in fact, is sufficient to
recover the quantum mechanical predictions for any joint measurement
of Pauli observables on this tripartite state. To recover the quantum
mechanical predictions associated with Pauli measurements on
$n$-partite GHZ states, and indeed on any $n$-partite state which can
be generated using only Gottesman-Knill operations, no more than $n-2$
bits are required (see: \citealt[]{tessier2004,tessier2005}; the
procedure is also sketched in Appendix \ref{sec:appendixTessier}).

It will be best to stop here for a moment, as the reader may, not
without some justification, feel uncomfortable at the prospect of
including any form of communication as part of a \emph{local} hidden
variables description. We have here two seemingly obvious and yet
prima facie incompatible claims. On the one hand it is clearly the
case that a classical computer is a classical physical system, whose
parts interact with one another in a locally causal, spatiotemporally
continuous, way. Thus a description of a classical computer
simulation of a quantum system, even when it involves communication,
is clearly a local hidden variables description of the quantum system
in some sense. On the other hand, it is also true that the supposed
nonlocal character of quantum systems is usually taken to be
demonstrated by the fact that certain quantum experiments seem to
require some form of influence (even if only benign) from one end of
the apparatus to the other in order to satisfy the Bell inequalities
\citep[see, e.g.,][]{maudlin2011}. There is a tension here that needs to
be resolved. Let us see if we can resolve it.\footnote{Presumably
  this tension is part of the reason why Tessier does not go so far
  as to call his scheme a local hidden variables description, but
  rather only a local hidden variables description that has been
  \emph{augmented} with classical communication. On Tessier's
  interpretation of the scheme it is not really either of these,
  however, for the imaginary terms in the scheme (see Appendix
  \ref{sec:appendixTessier}) evidently interfere nonlocally
  to preclude (in the absence of communication) an unambiguous
  assignment of spin components. The classical communication
  subsequently employed to compensate is thus not best viewed as a
  supplement to an otherwise local model but as a corrective to the
  nonlocal influences present in it. To turn the scheme into a local
  hidden variables description, one must reinterpret it in a way
  similar to the way depicted in Figure \ref{fig:localVnonlocal}, to
  be discussed in the next section.}

\section{Explicating the notion of a local hidden variables theory}
\label{sec:explic}

We can start to resolve the tension if we consider, first, just what
it is that we are required to reproduce. When a pair of spin-1/2
particles, $a$ and $b$, are prepared in the singlet state and then
subsequently spatially separated, then when Alice makes, for instance,
a $Z$ measurement on $a$, quantum mechanics tells us that the state of
$b$ changes instantaneously, so that when Bob measures $Z$ on $b$, his
result will be perfectly anti-correlated with Alice's. These are the
dynamics of quantum systems according to quantum mechanics. Our
challenge, however, is not to reproduce the dynamics of standard
quantum theory. Our task is to produce an \emph{alternative} theory
with the ability to reproduce the \emph{actually observed} statistics
associated with such experiments.\footnote{The local hidden
  variables account which I am about to describe is superficially
  different from, but essentially quite similar to what
  \citet[]{kent2005} has called the ``collapse locality loophole.''
  In Kent's loophole account, state reductions for Alice's and Bob's
  particles do not occur at spacelike but at timelike separation,
  while the precise character of the state reduction for any one
  particular particle depends on any previous state reductions
  within the past light cone of that particle. I am indebted to
  Wayne Myrvold for drawing my attention to Kent's account and to
  its similarity with the one sketched below.}

\begin{figure}
\begin{center}
  \includegraphics[scale=0.19]{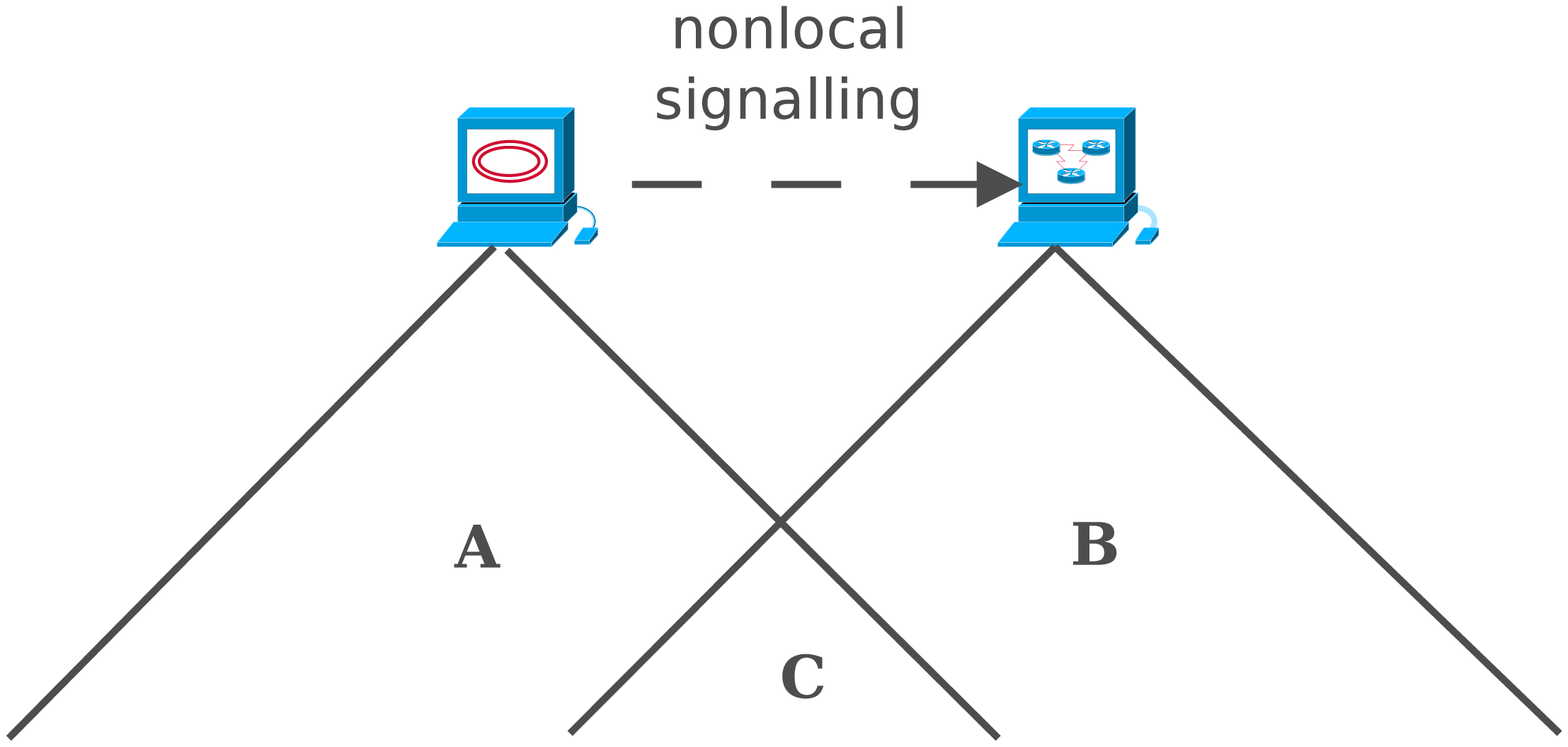}
  $\qquad$
  \includegraphics[scale=0.19]{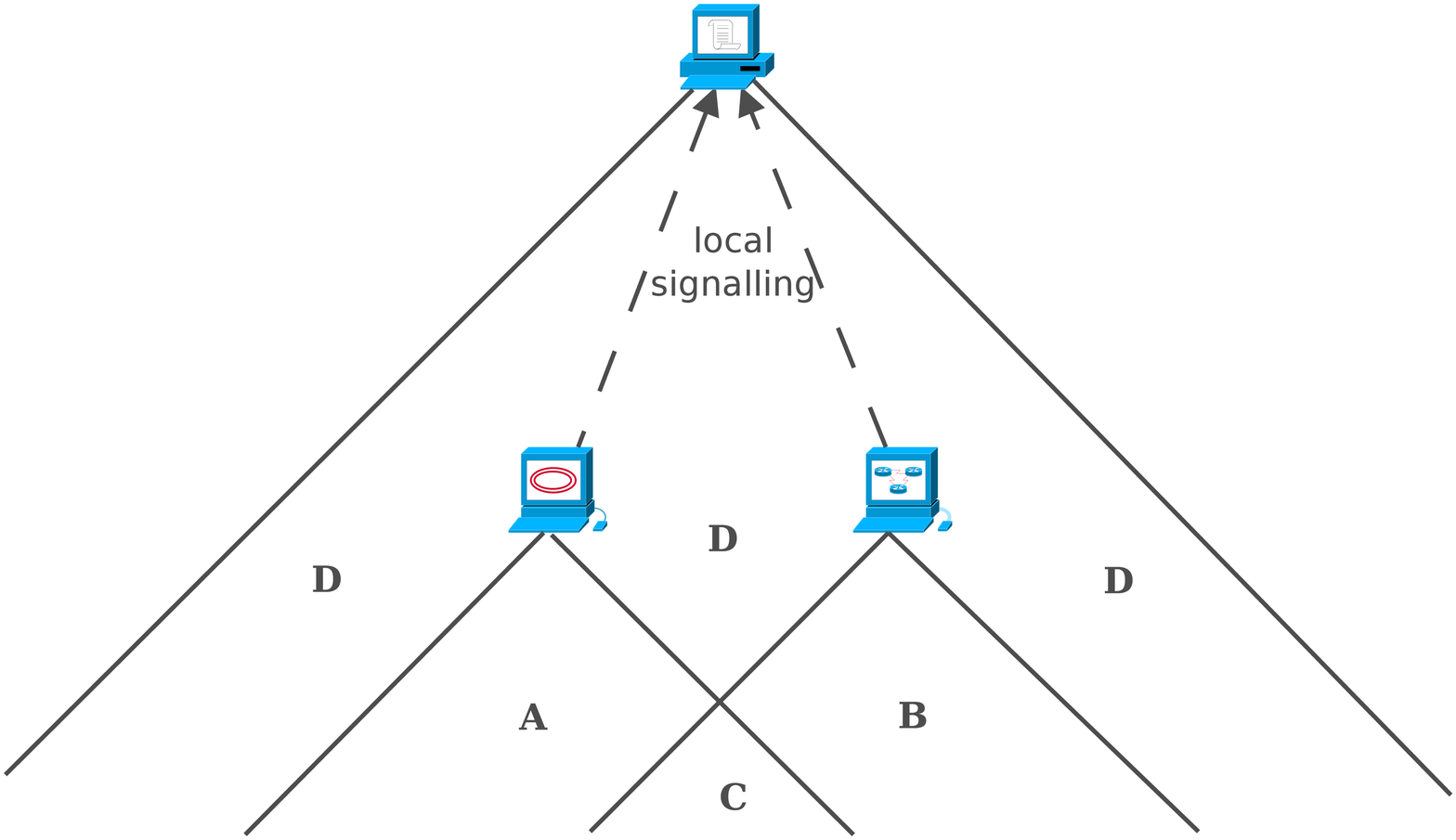}
\end{center}
\caption{Satisfying the Bell inequalities nonlocally (left) and locally
  (right).}
\label{fig:localVnonlocal}
\end{figure}

Now in order to actually observe the statistics associated with joint
Pauli experiments on two or more spatially separated systems, one must
combine the results associated with each of these
sub-experiments. That is, someone must somehow gather together the
results registered locally by Alice and Bob, place them side by side,
and examine them in order to observe the joint outcome. There is no
escaping this. Whether Alice and Bob physically meet with one another
to discuss the results over tea, or whether they physically transmit
their results to one another or to a neutral party via telephone, or
use some other physical means, it is absolutely necessary that the
results be collated together at some point, somehow, if the combined
outcome of the experiment is to be actually observed. During this
process of collating the results, there is time for Alice and Bob to
send finite signals to one another (at a velocity no greater than that
of light) so as to coordinate the observed outcomes of their
individual sub-measurements and ``correct'' them if necessary. If we
now take our measurement event to consist in the act of actually
observing the combined result, then all of this signalling activity
will have taken place in the past light cone of that measurement event
(see Figure \ref{fig:localVnonlocal}). It can therefore be considered
as part of the state preparation---the common cause---for that
measurement event. Thus if we produce a description of this sort which
replicates the statistics associated with a particular class of
observables, we will have produced an alternative local hidden
variables theory for those observables.

The tension I pointed out in the last section is still not completely
resolved. Surely, it will be objected, this is not what we normally
think of when we think of a local hidden variables theory. The device
of delaying the evaluation of the combined measurement result until
the parties have found time to meet together over crumpets and tea is
blatantly ad hoc. It is also extremely conspiratorial---quite
comically so. Yes, it most certainly is both, but this is no
fault in this context.

Bell's and related inequalities are best thought of, not as `no-go'
theorems per se, but rather as specifying \emph{constraints}. They
tell us what a local hidden variables theory which gives rise to a
joint probability distribution must be like; i.e., what general
properties it must satisfy. In and of themselves, however, Bell's and
related inequalities are little more than formal statements. Bell's
\citeyearpar{bell1964} original inequality, for instance, is
essentially little more than a theorem of probability
\citep[see][]{bell1981}.\footnote{It is true that we must appeal to
  a particular quantum state in order to show that it is
  \emph{violated} in some sense by quantum mechanics. But in and of
  itself it expresses only a general constraint that must be
  satisfied by factorisable probability distributions of a certain
  kind.} This is no fault. It is precisely this, in fact, which gives
it and similar inequalities their power and generality. Nevertheless,
if we are to make a meaningful distinction between what is and isn't
ruled out by them---if we are to take the step of turning them into
no-go theorems---then we must consider the context in which
they are being discussed.

Normally, there is no need to make the context of discussion explicit,
for normally it is just understood that we are situated in what I will
call the ``theoretical'' context. In this context, the Bell
inequalities are taken to inform our theories \emph{of the natural
  world}: they help us answer the question, specifically, of whether
there is some deeper underlying theory of the natural world in
relation to which quantum mechanics is only an approximation. They
cannot answer this question all by themselves, however. Any theory of
the natural world must do more than merely satisfy the constraints
imposed by the Bell inequalities. It must, in particular, also be
\emph{plausible}. Thus besides reproducing the statistical predictions
of quantum mechanics, it must also be consistent with our other
theories of physics (and if not it will need to provide a convincing
reason why those should be modified), and with the body of our
experiential knowledge in general. It is in fact on the basis of these
plausibility constraints, and not on the basis of the Bell
inequalities, that we rule out many of the ``loopholes'' left open by
the Aspect et. al. experiments.\footnote{Kent, for example, provides
  a number of strong (though for him, not conclusive) plausibility
  considerations of just this kind for ruling out his own ``collapse
  locality loophole'' \citep[6]{kent2005}.} Thus, on the basis of
these plausibility constraints, a local hidden variables theory in
which individual subsystems communicate classically with one another
(or with some ``neutral'' system), and in which the results of these
individual sub-experiments are only ever finalised once they have all
been collated together---however far in the future this might be from
the initial sub-experiments, and no matter what has happened to Alice
and Bob in the time between---is not only implausible, it is
completely absurd.

The theoretical context is not the only context in which the Bell
inequalities are relevant, however. There is also what I will call the
\emph{purely conceptual} context. In this context the Bell
inequalities are taken to inform us as to what is \emph{logically}
possible and still consistent with the predictions of quantum
mechanics. In this context there really are no, or at any rate no
fixed, plausibility constraints on alternative theories. This is the
realm of the ``toy theory.'' And yet this is still a useful context of
inquiry. It is useful, in particular, for making conceptual
distinctions. Thus Maudlin's \citeyearpar[89-90]{maudlin2011}
criticism of Howard's claim that outcome independence implies
separability, for instance, utilises a toy theory of this
sort.\footnote{In the (nonlocal) model, two identical particles,
  equipped with some means of superluminal communication (a tachyon,
  for instance), interact and then are sent in the direction of two
  detectors at spacelike separation from one another. Each particle is
  equipped with an identical set of instructions: on approaching a
  detector, if no message has been received, the particle is to
  effectively flip a fair coin to determine whether or not to pass
  through, after which it is to transmit a message (e.g., via the
  tachyon) specifying the setting of its detector and the outcome of
  the experiment (i.e., whether or not it passed through). Upon
  receiving this message, the other particle is to agree with the
  outcome specified in the message with a probability calculated on
  the basis of the difference between the setting of its detector
  and the detector setting specified in the message it has
  received.} To criticise Maudlin on account of the toy theory's
physical implausibility (see, e.g., \citealt[n. 33]{berkovitz2008} for
a criticism of this kind), then, is to miss the point. The purpose of
Maudlin's toy theory is not to give a physically plausible separable
model for outcome dependence, but rather to demonstrate that there is
a conceptual distinction to be made between outcome independence and
separability.

The context that is most appropriate to a discussion of the respective
characteristics of quantum and classical computers is neither the
theoretical context, which is too narrow, nor the purely conceptual
context, which is too broad. But there is yet a third context which
the Bell inequalities are relevant to, which is distinct from both the
theoretical and purely conceptual contexts described above. I will
call this the \emph{practical} context. In the practical context we
are concerned with what \emph{we} are capable of doing with the aim of
classically reproducing the statistical predictions of quantum
mechanics. In this context we are concerned, that is, with the
classical physical systems that are possible for us to \emph{build}
with the aim of reproducing these statistics.

Now clearly, it is completely irrelevant in this context whether our
description of such a system (which may or may not be characterisable
as a ``computer'' in Turing's sense) constitutes a viable alternative
theory of the natural world. And yet unlike the purely conceptual
context there is a set of fixed plausibility constraints applicable
here. What are these? Well for one, if the system is to be classical,
then it must be characterisable in accord with the laws and principles
of classical physics. It cannot, therefore, include any form of
superluminal communication. But further: we are finite beings, with
only finite resources at our disposal. The complexity involved in the
specification of such a system must therefore be \emph{tractable}. We
will thus rule out any system which uses an infinite number of
additional resources, or even only an amount that is exponential with
respect to the number of systems, as implausible in this
context. Local hidden variables descriptions which utilise only very
limited additional resources, on the other hand, will be perfectly
plausible.\footnote{I am taking ``tractable'' here in a relative
  sense. That is, the resources required by a classical computer to
  reproduce a particular effect should differ tractably from those
  required by a quantum computer. Or in other words: it should
  \emph{not be essentially harder} for the classical system to
  produce the same effect as the quantum system.}

The context most appropriate to a discussion of the respective
properties of quantum and classical computers is obviously the
practical, not the theoretical or purely conceptual contexts; for
quantum and classical computers are physical systems buildable by
\emph{us}. The local hidden variables description depicted in Figure
\ref{fig:localVnonlocal}, therefore, is a plausible local hidden
variables theory of just this, i.e. of the \emph{practical}, kind.

Before moving on, let me discuss an issue regarding the distinctions
that I am making which may have begun to worry the informed
reader. Earlier (\textsection \ref{sec:siggkthm}), during our
discussion of local hidden variables descriptions of bipartite
entangled states (i.e., the Bell states and other entangled bipartite
combinations of Pauli eigenstates), we saw that it was possible to
provide a local hidden variables description to reproduce the
statistics associated with Pauli measurements on systems in such
states, and moreover without utilising any amount of
communication. But in fact it is possible, if we allow the parties to
exchange a \emph{single} classical bit, to provide what I have been
calling a practical local hidden variables description of the
statistics for these states---not only for Pauli measurements---but
for \emph{any} combined projective measurement on the system (see
Appendix \ref{sec:appendixBellCc}). Given this, the reader may wonder
whether the practical context really rules out anything as
implausible; i.e., whether there are any meaningful distinctions
between local and nonlocal theories to be made in that
context.

Indeed, if it were the case that the statistics associated with
\emph{all} quantum mechanical measurements---regardless of the state
and the number of systems they were performed upon---could be
accounted for by a local hidden variables description of this kind,
then truly nothing would be ruled out as implausible in this
context. But that would nevertheless be a very meaningful claim. It
would mean, precisely, that it is both possible and plausible for us
to build classical physical systems to reproduce every observable
quantum mechanical effect. Surely this would be of enormous
interest. This is not the case, of course. But in the same way, it is
certainly both highly meaningful and of great interest to say that it
is both possible and plausible for us to build classical physical
systems to reproduce all of the observable behaviour of systems in the
Bell states.

At any rate it is actually not very surprising (in hindsight) that the
statistics associated with arbitrary projective measurements on
bipartite entangled systems are recoverable using only a small amount
of additional resources. Both \citet[]{jozsa2003} and
\citet[]{abbott2010} have shown,\footnote{For
  \citeauthor[]{jozsa2003}, of course, the proof is only taken as
  valid in the context of the amplitude formalism.} by independent
means, that one requires \emph{multi}-partite entanglement (i.e., a
number of parties $n \geq 3$) for a quantum computer to realise a
computational advantage over a classical computer. All three of these
results, therefore, are mutually supporting. But importantly,
regarding the multi-partite case: while there are strong indications
\citep[see][]{tessier2004} that the number of classical bits required
to model the quantum mechanical predictions associated with arbitrary
projective measurements is unbounded, the number of bits required to
classically reproduce the statistics associated with \emph{Pauli}
measurements is only linear in the number of systems. Thus we can
plausibly build---i.e., produce local hidden variables descriptions
of---classical systems to model the statistics associated with the
Gottesman-Knill operations for an arbitrary number of quantum
systems. But we cannot plausibly do so for arbitrary operations and
measurements.

Of course, in the \emph{theoretical} context, even one bit of
classical communication would be completely implausible. Indeed I will
venture to say that we would never take seriously any theory
involving communication of this kind which purports to be a theory of
the natural world. But when we are discussing the differences between
classical and quantum computers, we are situated not in the
theoretical but in the practical context: we are discussing the kinds
of things that \emph{we} can do and the kinds of systems that
\emph{we} can build.

\section{The sufficiency of entanglement thesis}
\label{sec:suff}

Let us review the ground we have covered so far. Recall that according
to the Gottesman-Knill theorem, there is a certain set of quantum
operations which, despite the fact that they are capable of generating
entangled states, are efficiently simulable on a classical
computer. And recall that based on this, it has been claimed that
entanglement is not, therefore, sufficient to enable one to achieve a
quantum speedup. Now we saw (\textsection \ref{sec:siggkthm}) that to
say of a quantum system that it is ``simulable by a classical
computer,'' is essentially to say that it is possible to give it an
alternative description in terms of local hidden variables, in a sense
that we clarified in \textsection \ref{sec:explic}. Thus the
Gottesman-Knill theorem entails the following: there are some
statistics associated with entangled states that admit of a local
hidden variables description.

We already knew this, however, and contrary to \citet[]{jozsa2003}, we
did not have to refer to the stabiliser formalism to learn it. For
Bell's and related inequalities inform us of the general constraints
that a locally causal description of a joint probability distribution
must satisfy. And in some cases (i.e., for some measurements), as we
already knew, the predictions associated with such descriptions are
compatible with the quantum mechanical ones. Moreover, we did not need
the Gottesman-Knill theorem to tell us that, as we saw in \textsection
\ref{sec:explic}, the local hidden variables descriptions compatible
with the statistics of Pauli measurements are such that we would
consider them plausible in the context of a discussion of classical
physical machines that we can build. Finally, as I will elaborate upon
shortly, we already know, and do not need the Gottesman-Knill theorem
to tell us, that this particular set of quantum operations is
minuscule as compared to the set of all operations performable on such
quantum systems.

It is possible to characterise the distinction between classical and
quantum mechanical systems as follows. Whereas the nature of quantum
mechanical systems is such that they allow us to efficiently exploit
the full representational capacity of Hilbert
space,\footnote{\citet[Ch. 8]{duwell2004} calls this feature
  `well-adaptedness'.} this is not so for classical systems
\citep[see][]{ekert1998}. The state space of an $n$-fold quantum
system that is efficiently simulable classically is only a tiny
portion of the system's overall state space. The reason for the larger
size of the quantum state space ($2^n$ dimensions for $n$ qubits), however, is the possibility of
entangled systems. It is because composite classical systems must
always be representable as product states that their state space is
smaller ($2n$ dimensions for $n$ bits). If we have an $n$-fold entangled quantum system, therefore,
it follows straightforwardly that the evolution of such a system
cannot, \emph{in general}, be efficiently simulated
classically.\footnote{I say `efficiently' because it is always
  possible, of course, to simulate quantum evolution if one allows for
  an exponential slowdown.} It is nevertheless possible, of course, to
utilise only a small portion of the state space of a quantum
mechanical system\textemdash exactly that portion of the state space
which is accessible efficiently by a similarly sized classical
system\textemdash but this has no bearing on the nature of the actual
physical resources that are provided by the quantum
system. Analogously, a life vest may be said to be sufficient to keep
me afloat on liquid water. I must actually wear it if it is to perform
this function, of course; but that is not a fact about this piece of
equipment's capabilities, only about my choice of whether to use it or
not.

It is therefore misleading, I believe, to conclude on the basis of the
Gottesman-Knill theorem, that entanglement is not a sufficient
resource to enable quantum computational speedup. What the
Gottesman-Knill theorem shows us--something we should already
know---is that simply having a system in an entangled state is not
enough to give one a quantum speedup. One must also \emph{use} such a
system to its full potential; i.e., one must not limit oneself to only
a minuscule proportion of the operations allowable on such
systems. Obviously, entanglement is insufficient in this, somewhat
trivial, sense. But if one intends (as \citeauthor{jozsa2003} clearly
do) by the claim that entanglement is insufficient---something very
different---that \emph{further physical resources} are
required to enable speedup, then I submit that this has not been shown
by the Gottesman-Knill theorem.\footnote{Note that my characterisation
  of entanglement as a physical resource is not motivated only by
  the conceptual arguments I have just given. To cite but one of
  many examples, one can show \citep[]{masanes2006} that for any
  non-separable state $\rho$, some other state $\sigma$ is capable
  of having its teleportation fidelity \citep[see][\textsection
  9.2.2]{nielsenChuang2000} enhanced by $\rho$'s presence. It is
  also possible to quantify the amount of entanglement contained in
  a given state by means of so-called entanglement measures, the
  theory of which is surveyed in \citet[]{plenio2007}. Conceptual
  considerations aside, these uses legitimate, in this author's
  mind, the characterisation of entanglement as a physical
  resource. I thank Jeffrey Barrett for raising this issue.} Far from
being a problem for the view that entanglement is a sufficient
resource to enable quantum speedup, in fact, the Gottesman-Knill
theorem serves to \emph{highlight} the role that is actually played by
entanglement in the quantum computer and to clarify exactly why and in
what sense it is sufficient to preclude the computer's evolution from
plausibly being classically
simulated.\footnote{\label{fn:roughwaves}What if the waves are
  rough? It may be that in this case my life vest will not be
  sufficient to save me. Analogously, in the presence of noise, as
  noted by \citet[]{linden2001}, entanglement may not be sufficient
  to enable one to achieve \emph{exponential} quantum
  speedup. Nevertheless, even in rough weather I will at least have
  a better chance of surviving with my life vest on than I will
  without it. Likewise, even in the presence of noise, it seems likely
  that a system in an entangled quantum state will be sufficient to
  enable some (though perhaps only a sub-exponential) quantum
  speedup. For further discussion, see \citet[]{cuffaro2013}.
}

\section{Summary}

I began by introducing the Gottesman-Knill theorem with a view to
motivating the assertion that the sufficiency of entanglement thesis
is false in some sense, and then considered a detailed statement of
this assertion due to \citet[]{jozsa2003}. I subsequently clarified
what I argued to be the real lesson of the Gottesman-Knill theorem:
that there some statistics associated with entangled states which
admit of a local hidden variables description. I then considered the
consequences of this for our understanding of the sufficiency of
entanglement thesis. I concluded that the Gottesman-Knill theorem does
not show that the sufficiency of entanglement thesis, interpreted as a
claim regarding the resources required to enable quantum speedup, is
false.

\appendix

\section{Recovering statistics for bipartite entangled states
  generated from Gottesman-Knill operations}

\subsection{No communication}
\label{sec:appendixBellNoCc}

Recall that the Gottesman-Knill operations consist of the Clifford
group of transformations (possibly conditioned on classical bits)
which map the Pauli group into itself, measurements of observables in
the Pauli group, and state preparation in the computational
basis. Consider, to start with, state preparation. By hypothesis, each
qubit will initially be prepared in one of the states $| 0 \rangle$ or
$| 1 \rangle$. These states are stabilised by $Z$ and $-Z$
respectively; i.e., each qubit will begin in a state equivalent to
either $Z| 0 \rangle$ or $-Z| 1 \rangle$. The Pauli gates $X$, $Y$,
and $Z$ ($I$ is just the trivial transformation) represent $\pi$
rotations of the Bloch sphere about the $x$, $y$, and $z$ axes
respectively. Applied to $Z$ they yield: $XZX^\dagger = -Z$,
$YZY^\dagger = -Z$, $ZZZ^\dagger = Z$. Applied to $X$ and $Y$ they
yield: $XXX^\dagger = X$, $YXY^\dagger = -X$, $ZXZ^\dagger = -X$,
$XYX^\dagger = -Y$, $YYY^\dagger = Y$, $ZYZ^\dagger = -Y$. The
Hadamard gate is a $\pi/2$-rotation about the $y$-axis, followed by a
$\pi$-rotation about $x$. Applied to $X$, $Y$, and $Z$ it yields
$HXH^\dagger = Z$, $HYH^\dagger = -Y$, and $HZH^\dagger = X$. The
Phase gate ($R$) is a $\pi/2$ rotation about the $z$-axis, with:
$RXR^\dagger = Y$, $RYR^\dagger = X$, $RZR^\dagger = Z$. The CNOT gate
is a two qubit gate but its result is either an $X$ or $I$
transformation applied to the target qubit.

If we now also take account of the way each of these operations
transform the system (i.e., $H| 0 \rangle = | + \rangle$, $R| +
\rangle = | y+\rangle$, and so on), then it is easy to verify that the
combined effect of any of these operations, for any subsystem of the
system, must be equivalent to the measurement of one of the Pauli
observables $\pm X$, $\pm Y$, $\pm Z$ on an eigenstate of a Pauli
observable. The reader can easily verify that this fact continues to
hold if we also include the generalisations of the Pauli operators
$\pm iX$, $\pm iY$, and $\pm iZ$ among our allowed operations. Thus
this fact holds true for all of the Gottesman-Knill operations.

One of the states generable from the Gottesman-Knill operations
\eqref{eqn:psiminus} is the singlet state. Expectation values for the
results of joint Pauli measurements on this state are: $\langle X
\otimes X \rangle = \langle Y \otimes Y \rangle = \langle Z \otimes Z
\rangle = -1$, with the expectation value for all other joint
measurements (of Pauli observables) $= 0$. A local hidden variables
theory to recover these (and only these) statistics was given by
\citet[]{bell1964}:\footnote{In Bell's original (equivalent)
  version, $B_{\hat{\lambda}}(\hat{n}) = -\mbox{sign}(\hat{n}\cdot
  \hat{\lambda}).$ The reason for my modified presentation will
  become evident shortly.}
\begin{align*}
A_{\hat{\lambda}}(\hat{m}) = \mbox{sign}(\hat{m}\cdot\hat{\lambda}), \\
B_{\hat{\lambda}}(\hat{n}) = \mbox{sign}(\hat{n}\cdot T(\hat{\lambda})).
\end{align*}
Here, $\hat{\lambda}$ is a local hidden variable, in the form of a
unit vector, taken on by both systems at the time of state
preparation. $\hat{m}$ and $\hat{n}$ are the measurement angles
associated with Alice's and Bob's experimental devices,
respectively. $A_{\hat{\lambda}}(\hat{m}), B_{\hat{\lambda}}(\hat{n})
\in \{\pm 1\}$ represent the results, given $\hat{\lambda}$, of spin
experiments on Alice's and Bob's subsystems. The function
sign($\alpha$) = 1 if $\alpha > 0$ or -1 if $\alpha < 0$. Note that in
Cartesian coordinates:
$$\hat{x} =
\left[\begin{smallmatrix} 1 \\ 0
    \\ 0 \end{smallmatrix}\right],
\hat{y} = \left[\begin{smallmatrix} 0 \\ 1
    \\ 0 \end{smallmatrix}\right],
\hat{z} = \left[\begin{smallmatrix} 0 \\ 0
    \\ 1 \end{smallmatrix}\right],
\hat{\lambda} = \left[\begin{smallmatrix} \cos\theta\sin\phi
    \\ \sin\theta\sin\phi \\ \cos\phi \end{smallmatrix}\right],$$
where $\theta$ is the angle $\hat{\lambda}$ makes with the $x$-axis in
the $x-y$ plane and $\phi$ is the angle $\hat{\lambda}$ makes with the
$z$-axis in the $z-x$ plane. Finally, $T$ is a transformation (some
combination of rotations and reflections) such
that $$T\left[\begin{smallmatrix} x \\ y
    \\ z \end{smallmatrix}\right] = \left[\begin{smallmatrix} -x
    \\ -y \\ -z \end{smallmatrix}\right].$$

Local hidden variables theories for Pauli measurement statistics on
the other Bell states, as well as on other bipartite entangled
combinations of Pauli eigenstates, can be obtained by varying $T$ (see
Figure \ref{fig:bellLikeLhvts}).

\begin{figure}
\begin{framed}
{
\footnotesize
\begin{tabular}{p{2cm} p{6cm} p{2.5cm}}
\underline{State} & \underline{Pauli Exp. values} & \underline{Transf.}
\\[10pt]
$| \Phi+ \rangle$ &
$\langle X \otimes X
\rangle = \langle Z \otimes Z \rangle = 1, \langle Y \otimes Y \rangle
= -1$ &
$T\left[\begin{smallmatrix} x \\ y \\ z \end{smallmatrix}\right] =
\left[\begin{smallmatrix} x \\ -y \\ z \end{smallmatrix}\right]$
\\[10pt]
$| \Phi- \rangle$
& $\langle Y \otimes Y \rangle = \langle Z \otimes Z \rangle = 1,
\langle X \otimes X \rangle = -1$ &
$T\left[\begin{smallmatrix} x \\ y
    \\ z \end{smallmatrix}\right] = \left[\begin{smallmatrix} -x \\ y
    \\ z \end{smallmatrix}\right]$
\\[10pt]
$| \Psi+ \rangle$ &
$\langle X \otimes X \rangle = \langle Y \otimes Y \rangle = 1,
\langle Z \otimes Z \rangle = -1$ &
$T\left[\begin{smallmatrix} x \\ y
    \\ z \end{smallmatrix}\right] = \left[\begin{smallmatrix} x \\ y
    \\ -z \end{smallmatrix}\right]$
\\[10pt]
$\frac{| 0,- \rangle - | 1,+ \rangle}{\sqrt 2}$ &
$\langle X \otimes Z \rangle = \langle Z \otimes X \rangle = -1,
\langle Y \otimes Y \rangle = 1$ &
$T\left[\begin{smallmatrix} x \\ y \\ z \end{smallmatrix}\right]
= \left[\begin{smallmatrix} -z \\ y \\ -x \end{smallmatrix}\right]$
\\[10pt]
$\frac{| 0,y- \rangle - | 1,y+ \rangle}{\sqrt 2}$ &
$\langle X \otimes Z \rangle = \langle Y \otimes X \rangle = \langle Z
\otimes Y \rangle = -1$ &
$T\left[\begin{smallmatrix} x \\ y \\ z \end{smallmatrix}\right]
= \left[\begin{smallmatrix} -y \\ -z \\ -x \end{smallmatrix}\right]$
\\[10pt]
$\frac{| +,y+ \rangle + | -,y- \rangle}{\sqrt 2}$ & $\langle X \otimes Y
\rangle = \langle Y \otimes X \rangle = \langle Z \otimes Z \rangle =
1$ &
$T\left[\begin{smallmatrix} x \\ y \\ z \end{smallmatrix}\right]
= \left[\begin{smallmatrix} y \\ x \\ z \end{smallmatrix}\right]$
\\[10pt]
\mbox{etc.}
\end{tabular}
}
\end{framed}
\caption{Examples of local hidden variables theories for bipartite
  entangled combinations of Pauli eigenstates. Expectation values for
  joint measurements of Pauli observables not indicated above are all
  $= 0$.}
\label{fig:bellLikeLhvts}
\end{figure}

\subsection{Communication}
\label{sec:appendixBellCc}

The following protocol is capable of reproducing the statistics
associated with arbitrary projections on spin-1/2 systems in the
singlet state \citep[]{toner2003}. Two independently chosen random
unit vectors, $\hat{\lambda}_1$ and $\hat{\lambda}_2$, are shared by
Alice and Bob at state preparation. Upon measuring her particle along
the direction $\hat{m}$, Alice outputs the result $A =
-\mbox{sign}(\hat{m} \cdot \hat{\lambda}_1).$ She then sends a
single classical bit $c = \mbox{sign}(\hat{m} \cdot\hat{\lambda}_1)
\times \mbox{sign}(\hat{m} \cdot\hat{\lambda}_2) = \pm 1$ to Bob. Bob,
who measures along $\hat{n}$, then outputs the result $B =
\mbox{sign}[\hat{n} \cdot (\hat{\lambda}_1 + c\hat{\lambda}_2)].$

\section{Recovering statistics of Pauli measurements on the GHZ state}
\label{sec:appendixTessier}

Below I sketch a procedure for consistently recovering the statistics
of Pauli observables on a tripartite GHZ state. For a more rigorous
treatment of this case, and for a treatment of $n$-partite GHZ states
and other states generable from Gottesman-Knill operations, see
\citet[]{tessier2004}.

Consider three spatially separated spin-1/2 systems, $a, b, c$, which,
having previously interacted, are now in the GHZ state:\footnote{The
  label ``GHZ'' actually refers not just to this but to the family
  of states: $\frac{1}{\sqrt 2}(| 0 \rangle^{\otimes n} \pm | 1
  \rangle^{\otimes n})$.}
\begin{align}
\frac{1}{\sqrt 2}(| 0 \rangle_a | 0 \rangle_b | 0 \rangle_c + | 1
\rangle_a | 1 \rangle_b | 1 \rangle_c).
\end{align}
It can be shown \citep[see][]{mermin1990} that any local hidden
variables theory (in which the parties are not allowed to communicate)
which attempts to assign noncontextual values to the $x$ and $y$
components of each particle's spin in this state will predict
that: $$v(X^a \otimes X^b \otimes X^c) = v(X^a \otimes I^b \otimes
I^c) \cdot v(I^a \otimes X^b \otimes I^c) \cdot v(I^a \otimes I^b
\otimes X^c) = -1,$$ where $v(M)$ is the result of measuring the
observable $M$. This contradicts the quantum mechanical prediction
of: $$\mbox{ }v(X^a \otimes X^b \otimes X^c) = 1.$$

The following is a scheme for reproducing the statistics of the GHZ
state:
\begin{align}
\label{eqn:tessierhvt}
\begin{matrix}
\\
\hline
X \\
Y \\
Z \\
I
\end{matrix}
\left|
\begin{matrix}
  q_A & q_B & q_C \\
  \hline
  R_2R_3     & R_2     &  R_3 \\
  iR_1R_2R_3 & iR_1R_2 &  iR_1R_3  \\
  R_1        & R_1     &  R_1 \\
  1          & 1       &  1
\end{matrix}
\right|
\begin{matrix}
\\
\hline
\mbox{ }\\
\mbox{ }\\
\mbox{ }\\
\mbox{ }.
\end{matrix}
\end{align}
Here, $q_A$, $q_B$, and $q_C$ are the constituent qubits of the
system. $X$, $Y$, $Z$, and $I$ refer to measurements of Pauli
observables. The $R_k$ are random variables which return a value of
$\pm 1$ with equal probability. They are to be interpreted
epistemically in the sense that they reflect our limited knowledge of
a determinate value of either $+1$ or $-1$ that is taken on by the
system at state preparation. Note that this value can only be revealed
by measurement: distinct systems subjected to identical state
preparations will in general have different values for their
$R_k$.\footnote{No such interpretation of the variables $R_k$ is
  given in either \citet[]{tessier2004} or \citet{tessier2005}, but
  such an interpretation is implicit if one is to make sense of the
  claims made therein.} To determine the outcome of a particular
measurement, we multiply the entries in the lookup table corresponding
to the sub-measurements performed on each qubit, with $(R_k)^2 = 1$,
discarding any lone straggling value of $i$ that remains after
calculating the final result. For example, $v(X \otimes X \otimes X) =
R_2R_3R_2R_3 = 1$, $v(X \otimes Y \otimes Y) = R_2R_3iR_1R_2iR_1R_3 =
-1$, $v(Y \otimes Y \otimes X) = iR_1R_2R_3iR_1R_2R_3 = -1$, $v(X
\otimes Y \otimes I) = R_2R_3iR_1R_2 = \pm i \Rightarrow \pm 1$,
etc. In Figure \ref{fig:tessiercomplete} hidden variables tables for
all of the intermediate steps from the initial preparation of the
tripartite product state $| 000 \rangle$ to the final GHZ state are
depicted.

\begin{figure}
\begin{framed}
{
\scriptsize
\begin{align*}
| 000 \rangle: \qquad
\begin{matrix}
\\
\hline
X \\
Y \\
Z \\
I
\end{matrix}
\left|
\begin{matrix}
  q_A & q_B & q_C \\
  \hline
  R_1     & R_2     &  R_3  \\
  -iR_1   & iR_2    &  iR_3 \\
  1       & 1       &  1 \\
  1       & 1       &  1
\end{matrix}
\right|
\begin{matrix}
\\
\hline
\mbox{ }\\
\mbox{ }\\
\mbox{ }\\
\mbox{ }
\end{matrix}
& \quad \xrightarrow{H_A} \quad
\begin{matrix}
\\
\hline
X \\
Y \\
Z \\
I
\end{matrix}
\left|
\begin{matrix}
  q_A & q_B & q_C \\
  \hline
  1     & R_2     &  R_3  \\
  iR_1  & iR_2    &  iR_3 \\
  R_1   & 1       &  1 \\
  1     & 1       &  1
\end{matrix}
\right|
\begin{matrix}
\\
\hline
\mbox{ }\\
\mbox{ }\\
\mbox{ }\\
\mbox{ }
\end{matrix} \\
\xrightarrow{\mbox{\scriptsize{CNOT}}_{AB}} \quad
\begin{matrix}
\\
\hline
X \\
Y \\
Z \\
I
\end{matrix}
\left|
\begin{matrix}
  q_A & q_B & q_C \\
  \hline
  R_2     & R_2     &  R_3  \\
  iR_1R_2 & iR_1R_2 &  iR_3 \\
  R_1     & R_1     &  1 \\
  1       & 1       &  1
\end{matrix}
\right|
\begin{matrix}
\\
\hline
\mbox{ }\\
\mbox{ }\\
\mbox{ }\\
\mbox{ }
\end{matrix}
& \quad \xrightarrow{\mbox{\scriptsize{CNOT}}_{AC}} \quad
\begin{matrix}
\\
\hline
X \\
Y \\
Z \\
I
\end{matrix}
\left|
\begin{matrix}
  q_A & q_B & q_C \\
  \hline
  R_2R_3     & R_2     &  R_3 \\
  iR_1R_2R_3 & iR_1R_2 &  iR_1R_3  \\
  R_1        & R_1     &  R_1 \\
  1          & 1       &  1
\end{matrix}
\right|
\begin{matrix}
\\
\hline
\mbox{ }\\
\mbox{ }\\
\mbox{ }\\
\mbox{ }
\end{matrix}
\end{align*}
}
\end{framed}
\caption{A series of hidden variables tables modelling the preparation
  of the state GHZ $= (| 000 \rangle + | 111 \rangle) / \sqrt 2$
  \citep[]{tessier2004}. The update rules for the H and CNOT gates
  are: \textbf{H:} $X^f = Z^i$, $Y^f = -Y^i$, $Z^f = X^i$.
  \textbf{CNOT:} $X_s^f = X_s^iX_t^i$, $Y_s^f = Y_s^iX_t^i$,
  $Z_s^f = Z_s^i$, $X_t^f = X_t^i$, $Y_t^f = Z_s^iY_t^i$, $Z_t^f =
  Z_s^iZ_t^i$, where $P^i$ is the specification of $P$ before the
  given transformation and $P^f$ is its new specification. $s$ and
  $t$ refer to the control and target qubits, respectively, involved
  in a given CNOT operation.%
}
\label{fig:tessiercomplete}
\end{figure}

It can be verified that all of the predictions of quantum mechanics
for joint Pauli experiments on the GHZ state are recovered by this
scheme. Unfortunately the results of these experiments cannot be made
consistent with one another under the assumption that each qubit's $x$
and $y$ spin component is a noncontextual ``element of reality''
associated with the system. For instance, the outcome of the joint
measurement $X \otimes Y \otimes Y$ is $v(X \otimes Y \otimes Y) =
R_2R_3iR_1R_2iR_1R_3 = -1$. Under the supposition that each qubit
possesses independent values of both spin-$x$ and spin-$y$, this joint
measurement outcome must be consistent with the product of the
outcomes of the individual measurements $X \otimes I \otimes I$, $I
\otimes Y \otimes I$, and $I \otimes I \otimes Y$. But
it is not, for $v(X \otimes I \otimes I) \times v(I \otimes Y \otimes
I) \times v(I \otimes I \otimes Y) = (R_2R_3)(R_1R_2)(R_1R_3) = 1$.

This can, however, be compensated for. We can ensure consistency by
allowing the parties to signal. For instance, Bob and Alice can agree
that he will send her a single classical bit indicating whether or not
he performed a $Y$ measurement on his qubit. Upon receipt of this bit,
Alice should flip the sign of her local outcome if either she or Bob
(or if both of them) measured $Y$. Thus for the case above we will
have $v(X \otimes I \otimes I) \times v(I \otimes Y \otimes I) \times
v(I \otimes I \otimes Y) = (-R_2R_3)(R_1R_2)(R_1R_3) = -1$. This is
consistent both with the value obtained for the joint measurement $X
\otimes Y \otimes Y$ and with the individual results for the
measurements $X \otimes I \otimes I$, $I \otimes Y \otimes I$, and $I
\otimes I \otimes Y$ (each of the latter three produces a random
outcome of $\pm 1$ with equal probability). In this manner it is
possible to make the outcome of every joint measurement specifiable in
the model consistent with the corresponding product of individual
measurement outcomes.

The scheme generalises. For an $n$-qubit GHZ state, as well as for any
state generable using only Gottesman-Knill operations, only $n-2$ bits
of classical communication are required to accurately model the
statistics associated with measurements of Pauli observables on the
system (details are given in \citealt[]{tessier2004}).

\bibliographystyle{apa-good}
\bibliography{Bibliography}{}

\begin{thebibliography}{28}
\expandafter\ifx\csname natexlab\endcsname\relax\def\natexlab#1{#1}\fi
\expandafter\ifx\csname url\endcsname\relax
  \def\url#1{{\tt #1}}\fi
\expandafter\ifx\csname urlprefix\endcsname\relax\def\urlprefix{URL }\fi

\bibitem[{Aaronson(2013)}]{aaronson2013}
Aaronson, S. (2013).
\newblock {\em Quantum Computing Since {D}emocritus\/}.
\newblock New York: Cambridge University Press.

\bibitem[{Abbott(2012)}]{abbott2010}
Abbott, A.~A. (2012).
\newblock The {Deutsch-Jozsa} problem: De-quantisation and entanglement.
\newblock {\em Natural Computing\/}, {\em 11\/}, 3--11.

\bibitem[{Bell(2004 {[1964]})}]{bell1964}
Bell, J.~S. (2004 {[1964]}).
\newblock On the {Einstein-Podolsky-Rosen} paradox.
\newblock In {\em Speakable and Unspeakable in Quantum Mechanics\/}, (pp.
  14--21). Cambridge: Cambridge University Press.

\bibitem[{Bell(2004 {[1981]})}]{bell1981}
Bell, J.~S. (2004 {[1981]}).
\newblock Bertlmann's socks and the nature of reality.
\newblock In {\em Speakable and Unspeakable in Quantum Mechanics\/}, (pp.
  139--158). Cambridge: Cambridge University Press.

\bibitem[{Berkovitz(2008)}]{berkovitz2008}
Berkovitz, J. (2008).
\newblock Action at a distance in quantum mechanics.
\newblock In E.~N. Zalta (Ed.) {\em The Stanford Encyclopedia of Philosophy\/}.
  Winter 2008 ed.

\bibitem[{Cuffaro(2013)}]{cuffaro2013}
Cuffaro, M.~E. (2013).
\newblock {\em On The Physical Explanation for Quantum Computational
  Speedup\/}.
\newblock Ph.D. thesis, The University of Western Ontario, London, Ontario.

\bibitem[{Datta et~al.(2005)Datta, Flammia, \& Caves}]{datta2005}
Datta, A., Flammia, S.~T., \& Caves, C.~M. (2005).
\newblock Entanglement and the power of one qubit.
\newblock {\em Physical Review A\/}, {\em 72\/}, 042316.

\bibitem[{Deutsch(1997)}]{deutsch1997}
Deutsch, D. (1997).
\newblock {\em The Fabric of Reality\/}.
\newblock New York: Penguin.

\bibitem[{Deutsch et~al.(2000)Deutsch, Ekert, \& Lupacchini}]{deutsch2000}
Deutsch, D., Ekert, A., \& Lupacchini, R. (2000).
\newblock Machines, logic and quantum physics.
\newblock {\em The Bulletin of Symbolic Logic\/}, {\em 6\/}, 265--283.

\bibitem[{Duwell(2004)}]{duwell2004}
Duwell, A. (2004).
\newblock {\em How to Teach an Old Dog New Tricks: Quantum Information, Quantum
  Computing, and the Philosophy of Physics\/}.
\newblock Ph.D. thesis, University of Pittsburgh, Pittsburgh.

\bibitem[{Ekert \& Jozsa(1998)}]{ekert1998}
Ekert, A., \& Jozsa, R. (1998).
\newblock Quantum algorithms: Entanglement-enhanced information processing.
\newblock {\em Philosophical Transactions of the Royal Society A\/}, {\em
  356\/}, 1769--1782.

\bibitem[{Gottesman(1999)}]{gottesman1999}
Gottesman, D. (1999).
\newblock The {H}eisenberg representation of quantum computers.
\newblock In S.~P. Corney, R.~Delbourgo, \& P.~D. Jarvis (Eds.) {\em Group22:
  Proceedings of the {XXII} International Colloquium on Group Theoretical
  Methods in Physics\/}, (pp. 32--43). Cambridge, MA: International Press.
\newblock Longer version available at: {arXiv:quant-ph/9807006v1}.

\bibitem[{Hagar(2007)}]{hagar2007b}
Hagar, A. (2007).
\newblock Quantum algorithms: Philosophical lessons.
\newblock {\em Minds \& Machines\/}, {\em 17\/}, 233--247.

\bibitem[{Hameroff(1998)}]{hameroff1998}
Hameroff, S. (1998).
\newblock Quantum computation in brain microtubules? the {P}enrose-{H}ameroff
  `{Orch OR}' model of consciousness.
\newblock {\em Philosophical Transactions of the Royal Society A\/}, {\em
  356\/}, 1869--1896.

\bibitem[{Jozsa \& Linden(2003)}]{jozsa2003}
Jozsa, R., \& Linden, N. (2003).
\newblock On the role of entanglement in quantum-computational speed-up.
\newblock {\em Proceedings of the Royal Society of London. Series A.
  Mathematical, Physical and Engineering Sciences\/}, {\em 459\/}, 2011--2032.

\bibitem[{Kent(2005)}]{kent2005}
Kent, A. (2005).
\newblock Causal quantum theory and the collapse locality loophole.
\newblock {\em Physical Review A\/}, {\em 72\/}, 012107.

\bibitem[{Linden \& Popescu(2001)}]{linden2001}
Linden, N., \& Popescu, S. (2001).
\newblock Good dynamics versus bad kinematics: Is entanglement needed for
  quantum computation?
\newblock {\em Physical Review Letters\/}, {\em 87\/}, 047901.

\bibitem[{Masanes(2006)}]{masanes2006}
Masanes, L. (2006).
\newblock All bipartite entangled states are useful for information processing.
\newblock {\em Physical Review Letters\/}, {\em 96\/}, 150501.

\bibitem[{Maudlin(2011)}]{maudlin2011}
Maudlin, T. (2011).
\newblock {\em Quantum Non-Locality and Relativity\/}.
\newblock Cambridge, MA: Wiley-Blackwell, third ed.

\bibitem[{Mermin(1990)}]{mermin1990}
Mermin, N.~D. (1990).
\newblock What's wrong with these elements of reality?
\newblock {\em Physics Today\/}, {\em 43\/}, 9--11.

\bibitem[{Nielsen \& Chuang(2000)}]{nielsenChuang2000}
Nielsen, M.~A., \& Chuang, I.~L. (2000).
\newblock {\em Quantum Computation and Quantum Information\/}.
\newblock Cambridge: {Cambridge University Press}.

\bibitem[{Plenio \& Virmani(2007)}]{plenio2007}
Plenio, M.~B., \& Virmani, S. (2007).
\newblock An introduction to entanglement measures.
\newblock {\em Quantum Information \& Computation\/}, {\em 7\/}, 1--51.

\bibitem[{Steane(2003)}]{steane2003}
Steane, A.~M. (2003).
\newblock A quantum computer only needs one universe.
\newblock {\em Studies in History and Philosophy of Modern Physics\/}, {\em
  34\/}, 469--478.

\bibitem[{Tessier(2004)}]{tessier2004}
Tessier, T.~E. (2004).
\newblock {\em Complementarity and Entanglement in Quantum Information
  Theory\/}.
\newblock Ph.D. thesis, The University of New Mexico, Albuquerque, New Mexico.

\bibitem[{Tessier et~al.(2005)Tessier, Caves, Deutsch, \& Eastin}]{tessier2005}
Tessier, T.~E., Caves, C.~M., Deutsch, I.~H., \& Eastin, B. (2005).
\newblock Optimal classical-communication-assisted local model of $n$-qubit
  {Greenberger-Horne-Zeilinger} correlations.
\newblock {\em Physical Review A\/}, {\em 72\/}, 032305.

\bibitem[{Timpson(2013)}]{timpson2013}
Timpson, C.~G. (2013).
\newblock {\em Quantum Information Theory \& the Foundations of Quantum
  Mechanics\/}.
\newblock Oxford: Oxford University Press.

\bibitem[{Toner \& Bacon(2003)}]{toner2003}
Toner, B.~F., \& Bacon, D. (2003).
\newblock Communication cost of simulating {B}ell correlations.
\newblock {\em Physical Review Letters\/}, {\em 91\/}, 187904.

\bibitem[{Wallace(2012)}]{wallace2012}
Wallace, D. (2012).
\newblock {\em The Emergent Multiverse\/}.
\newblock Oxford: Oxford University Press.

\end{thebibliography}

\end{document}